\documentclass[aps,prl, twocolumn]{revtex4}
\usepackage[T1]{fontenc}
\usepackage{framed}
\usepackage{tikz}
\usepackage{graphicx}
\usepackage{indentfirst}
\usepackage{latexsym}
\usepackage{multirow}
\usepackage{tabls}
\usepackage{color}
\usepackage{subfigure}

\usepackage{amsfonts}
\usepackage{amsmath}
\usepackage{bm}
\usepackage{mathrsfs}

\def\b{ \beta }

\def\b0{ {\bf 0} }

\newcommand{\be}{\begin{equation}}
\newcommand{\ee}{\end{equation}}

\def\lsim{\mathrel{\rlap{\lower3pt\hbox{\hskip1pt$\sim$}}
    \raise1pt\hbox{$<$}}}                
\def\gsim{\mathrel{\rlap{\lower3pt\hbox{\hskip1pt$\sim$}}
    \raise1pt\hbox{$>$}}}         

\def\coordeq{ \, \mathrel{ \rlap{\hbox{\hskip-2.5pt$=$} }
    \raise4pt\hbox{$\cdot$}} \, }                

\begin{document}

\title{Reply to ``Comment on `Finite size corrections to the radiation reaction force in classical electrodynamics' ''}

\def\addCMU{Department of Physics, Carnegie Mellon University, Pittsburgh PA  15213, USA}
\def\addMaryland{Maryland Center for Fundamental Physics, Department of Physics,
University of Maryland, College Park, MD 20742 USA}
\def\addModeling{Center for Scientific Computation and Mathematical
Modeling, University of Maryland, College Park, MD 20742 USA}
\def\addPitt{Department of Physics and Astronomy, University of
  Pittsburgh, PA 15260, USA} 

\def\addJPL{Jet Propulsion Laboratory, California Institute of Technology, Pasadena, California 91109, USA}
\def\addCaltech{Theoretical Astrophysics, California Institute of Technology, Pasadena, California 91125, USA}
\def\addUPitt{Pittsburgh Particle physics Astrophysics and Cosmology Center (PITT PACC)\\Department of Physics and Astronomy, University of Pittsburgh, Pittsburgh, Pennsylvania 15260, USA}

\author{Chad R. Galley}
\affiliation{\addJPL}\affiliation{\addCaltech}
\author{Adam K. Leibovich}
\affiliation{\addUPitt} 
\author{Ira Z. Rothstein}
\affiliation{\addCMU}


\maketitle

The authors of the ``Comment on `Finite size corrections to the radiation reaction force in classical electrodynamics'" \cite{comment} use the results of Nodvik \cite{Nodvik} to argue that a term is missed in the worldline action that yields an order $R$ correction to the Abraham-Lorentz-Dirac (ALD) formula.  We believe that the arguments made in the comment are correct. The missed term linear in $R$ in the worldline action is
\begin{align}
	C \int d\tau \, a^2
\label{eq:newterm}
\end{align}
where $a^2 = a^\mu a_\mu$ is the square of the four-acceleration and $C$ is an undetermined coefficient that is found by matching onto a calculation from the full theory. For the case of a spherical shell of charge, the authors of \cite{comment} show that $C = - (2/9) e^2 R$. Importantly, the term in (\ref{eq:newterm}) vanishes for a neutral body.

It is interesting to note, however, that this term (\ref{eq:newterm}), while formally of order $R$, will be suppressed in any laboratory experiment. Thus its effects on proposals to measure finite size effects will be negligible. To see this we recall that the worldline of the point charge in the effective theory is not an observable. Hence, one may perform worldline shifts by amounts of order $R$ without changing the predictions of the theory. In this regard, note that (\ref{eq:newterm}) can be eliminated by the following shift in the worldline 
\begin{align}
	x^\mu \rightarrow x^\mu - \frac{C}{m} \, a^\mu
\label{eq:shift}
\end{align} 
where $C$ is the coefficient of the $a^2$ term in (\ref{eq:newterm}) and scales as $e^2 R$ \cite{comment}. The result of this shift is that the original action \cite{prl}
\begin{align}
	S [ x^\mu, A_\mu ] = {} & -m \int d\tau + e \int d\tau \, u^\alpha A_\alpha (x) + C \int d\tau \, a^2 \nonumber \\
		& + C_d \int d\tau \, u^{[\alpha } a^{\beta ] } F_{\alpha \beta} (x) + \cdots
\label{eq:action1}
\end{align}
becomes
\begin{align}
	S[ x^\mu, A_\mu ] \rightarrow {} & - m \int d\tau + e\int d\tau \, u^\alpha A_\alpha (x) \nonumber \\
		& + \left(  \frac{ e C}{m } + C_d \right) \int d\tau \, u^{[\alpha } a^{\beta ] } F_{\alpha \beta} (x) + \cdots , \nonumber
\end{align}
which is obtained by expanding (\ref{eq:action1}) in powers of $R$ after applying the shift in (\ref{eq:shift}). There are two key points to note. The first is that (\ref{eq:newterm}) is {\it cancelled} from the action. The second is that the acceleration-induced dipole moment term that we studied in \cite{prl} (i.e., the last term in (\ref{eq:action1})) receives a correction to its matching coefficient. Hence, under this worldline shift, $C_d \rightarrow C_d + eC/m$.

Next, we ask under what conditions $e C/m$ is much smaller than $C_d$. Since $C \sim e^2 R$ \cite{comment} and $C_d \sim e R^2$ \cite{prl} it follows that $(e C / m ) / ( C_d ) \sim  e^2 /( m R)$, which is much less than unity if
\begin{align}
	m \gg \frac{ e^2 }{R }  .
\label{eq:inequality}
\end{align}
On the left side of (\ref{eq:inequality}) is the rest mass energy of the object and $e^2 / R$ is the potential energy of the extended charge. As the rest mass energy will almost assuredly be much larger than the charge's electrostatic energy in any real-world laboratory experiment then it follows that the contribution from (\ref{eq:newterm}) will not manifest itself. In fact, if one were in the situation where $m \sim e^2 / R$ then the electrostatic energy is comparable to the object's rest mass energy. In this case, quantum processes will become very important since the spontaneous production of identical charged objects from the vacuum will occur. Of course, in such a scenario, the effective theory of \cite{prl}, as well as any classical description of the extended charge, is no longer valid. Copyright 2012. All rights reserved.


\begin{thebibliography}{99}

\bibitem{comment} P. Forg\'acs, T. Herpay, P. Kov\'acs, ({\it Preprint} 1202.6289)

\bibitem{Nodvik} J. S. Nodvik, Ann. Phys. (NY) {\bf 28}, 225 (1964)

\bibitem{prl} C. R. Galley, A. K. Leibovich, and I. Z. Rothstein, Phys. Rev. Lett. {\bf 105}, 094802 (2010)

\end{thebibliography}
\end{document}